\documentclass{aastex}          
\usepackage{spr-astr-addons}
\usepackage{amsfonts}
\usepackage{amsmath}
\usepackage{latexsym}
\usepackage{url}
\usepackage{graphicx}[12pt]
\newcommand{\bm}[1]{\mbox{\boldmath$#1$}}

\DeclareMathOperator{\tg}{tg}

\newcommand{\bea}{\begin{eqnarray}}
\newcommand{\eea}{\end{eqnarray}}
\newcommand{\nn}{\nonumber}

\begin{document}


\title{Effective potential energy in St\o rmer's problem for an inclined rotating magnetic dipole}

\shorttitle{Effective potential energy}

\author{V. Epp
}

\and
\author{M. A. Masterova
}
\affil{Tomsk State Pedagogical University, ul. Kievskaya 60, 634061 Tomsk, Russia}
\email{epp@tspu.edu.ru}

\begin{abstract}
We discuss the dynamics of a charged nonrelativistic particle in electromagnetic field of a rotating magnetized celestial body. The equations of motion of the particle are obtained and some particular solutions are found.  Effective potential energy is defined on the base of the first constant of motion. Regions accessible and inaccessible for a  charged particle motion are studied and depicted for different values of a constant of motion.
\end{abstract}
\section{Introduction}

The field of magnetic dipole  and motion of charged particles in this field have large practical significance for astrophysics.
Magnetic field of many planets and stars can be thought of as a dipole field in good approximation. Stationary case, when the magnetic moment of a celestial body coincides with the rotation axis, is  well studied.
Particularly, motion of a charged particle in the Earth magnetic field is studied in detail. Solution of equation of motion for charged particle in a dipolar  magnetic field gives trapping regions for particles,
which have energy of limited range. These regions of  planets are named radiation belts \citep{Alfven,Holmes}.
The first theoretical studies on the properties of the trajectories of a charged
particle in the dipolar field where done by \citet{Stormer}, \citet{DeVogelaere} and \citet{Dragt}.
Rather complicated mathematical methods were developed in order to calculate the trajectory of charged particle in the dipole magnetic field (see for example \citep{Lanzano}). Solution of the St\o rmer's problem was extended to magnetic field that is a superposition of the field of a dipole and a uniform magnetic field. Allowed and forbidden regions of the motion of charged particles in such field was studied by \citet{Katsiaris}.

  There are also well known celestial  bodies,  which direction of magnetic moment differs from the direction of axis of rotation.
 In this case electric field is induced in the neighbourhood of the body by magnetic field. The neutron stars and pulsars are examples of such bodies. The first model of electric field
 which is generated  in the neighbourhood of a neutron star was developed by \citet{Deutsch}. Some other models were suggested and studied by several authors \citep{Michel}. All these models are based on suggestion that the neutron star is a conducting body.
  
Conducting body rotating in its own magnetic field, generates a potential difference between the poles and equator. This potential difference leads to development of a co-rotating  magnetosphere. By assuming that the magnetic field is dipolar, and unaffected by the trapped particles in the magnetosphere, and that the field dipole axis is parallel to the rotation axis, \citet{Gold} determined many of the properties of the magnetosphere. Electromagnetic field and magnetosphere of the oblique rotator and conducting body have been studied by many authors. See for example  \citet{Cohen, Kaburaki, Cohen80}.

Electromagnetic field of conducting body differs essentially from the
  pure dipole field. But there are also celestial bodies, which consist of non-conducting matter and their axis of rotation  is inclined with respect to the magnetic field axis.
  Magnetic field of such objects  in good approximation can be described as the field of an inclined rotating magnetic moment or "oblique rotator"  \citep{Babcock}.
Eelectromagnetic field in the near and intermediate zones of a magnetic dipole rotating in free space has been studied recently by \citet{Sarych}. He has also estimated the energy, which can be acquired by the particles during acceleration. 
 
In our previous paper \citep{EppMast} we have calculated the field of rotating non-conducting body
with the dipole magnetic field,  and  the dipole axis  inclined to the rotation axis.  An integral of motion was found and appropriate effective potential energy was studied. It was shown that each stationary point of the effective potential energy corresponds to particular solution of equations of motion for a charged particle. In this paper we develop investigation of the effective potential energy. We study here stability of the particle motion in the vicinity of the stationary points. Special attention is paid to investigation of regions accessible and inaccessible for a  charged particle motion. These regions are described and depicted for particles with different values of the constant of motion. It is shown that there are closed allowed regions at certain values of the constant of motion. These regions are co-rotating with the electromagnetic field of the magnetized body.

\section{The field of precessing magnetic dipole}
Let us consider the field which is produced by precessing magnetic dipole. We define the  law of motion of a dipole moment  $\bm\mu$   in the Cartesian coordinate system as follows:
\begin{eqnarray}
\label{eq1}
\bm\mu=\mu(\sin\alpha\cos\omega t,\, \sin\alpha\sin\omega t,\,\cos\alpha),
\end{eqnarray}
where $\mu>0$ and $\omega>0$ are the magnitude and angular velocity of the dipole, respectively, $0\leq\alpha\leq\pi$ is the angle between the
vector $\bm\mu$ and rotational axis.
Consider the general formulas for the field of an arbitrary changing dipole \citep{Feynman}:
\begin{equation}
\label{1a}
{\bm E}=\frac{(\bm {n}\times\dot {\bm {\mu}})}{r^2c}+\frac{(\bm {n}\times\ddot {\bm
{\mu}})}{rc^2},
\end{equation}
\begin{equation}
\label{1b}
\bm {H}=\displaystyle\frac{(\bm {n}\times(\bm {n}\times\ddot {\bm {\mu}}))}{rc^2}+\displaystyle\frac{3\bm
{n}(\bm {n}\dot {\bm {\mu}})-\dot {\bm {\mu}}}{r^2c}+\displaystyle\frac{3\bm {n}(\bm {n}\bm
{\mu})-\bm {\mu}}{r^3} ,
\end{equation}
where  $c$ is the speed of light, $\bm n=\bm r/r$ is the unit vector, $\bm r$ is the radius-vector.
Field is calculated at time $t$, and all the quantities on the right side of these equations are calculated at the retarded time ${t}'=t-\dfrac{r}{c}$.

We use a spherical coordinate system  ($r, \theta, \varphi$), in which
the components of electric field vector  have the form:
\begin{eqnarray}
\label{1}
 r^3E_r&=&0, \\
\label{2}
 r^3E_\theta&=&\mu\rho\sin\alpha \left(\rho\sin\tau-\cos\tau\right), \\
\label{3}
 r^3E_\varphi&=&-\mu\rho \cos \theta \sin\alpha \left( \rho\cos\tau+\sin\tau \right),
\end{eqnarray}
where
\begin{equation}
\label{3a}
\tau=\omega t'-\varphi,
\quad
\rho=\displaystyle\frac{r\omega}{c}.
\quad
\end{equation}
The magnetic field vector has components:
\begin{eqnarray}
\label{eq80}
r^3H_r&=&2\mu \left [ \sin \alpha \sin \theta (\cos
\tau-\rho \sin\tau)\right.\nonumber \\
&&\left.+\cos \theta \cos \alpha\right ],\\
\label{eq78}
 r^3H_\theta&=&-\mu\left [\cos \theta \sin \alpha (\cos\tau-\rho \sin\tau-\rho ^2\cos\tau)\right.\nonumber \\
& &\left.-\sin \theta \cos \alpha \right ],\\
\label{eq79}
r^3H_\phi &=&-\mu \sin \alpha (\sin\tau+\rho \cos\tau-\rho ^2\sin\tau )\, .
\end{eqnarray}
The same formulas for $\alpha=\pi/2$ have been obtained by \citet{Sarych}.  It follows from  Eqs  (\ref{1}) -- (\ref{3}) that the components of electric field satisfy equation:
\begin{equation}
\label{eq64}
\begin{array}{l}
\left( {\displaystyle\frac{E_{\theta}}{E_{0  \theta}}} \right)^2+\left( {\displaystyle\frac{E_{\varphi}}{E_{0  \phi}}} \right)^2=1,
 \quad
 \end{array}
\end{equation}
 where $E_{0  \theta}=\displaystyle\frac{\rho\mu\sin\alpha\sqrt{1+\rho^2}}{r^3}$, $E_{0 \varphi}= E_{0  \theta}\cos\theta$.
Hence, the vector $ \bm{E}$ circumscribes an ellipse with  semiaxises $E_{0  \theta}$ and $E_{0 \varphi}$ in the plane orthogonal to the  radius vector. This ellipse degenerates to a circle in the direction of axis of rotation ($\theta=0,\,\pi$), but in equatorial plane  $\theta=\displaystyle\frac{\pi}{2}$ vector $ \bm{E}$ oscillates in meridian plane (parallel to $z$-axis).

It follows from Eqs  (\ref{eq80}) and (\ref{eq78}) that the magnetic field has constant components (in contrast to the electric field):
\begin{equation}
H_{rc}=\displaystyle\frac{2\mu}{r^{3}}\cos\theta\cos\alpha,\quad
H_{\theta c}=\displaystyle\frac{\mu}{r^{3}}\sin\theta\cos\alpha.\nn
\end{equation}
These components relate to the field of a static magnetic dipole if the inclination angle $\alpha$ is equal to zero. The electric field tends to zero in this case.

One can see that time dependence  appears in the formulas for electric and magnetic fields  as composition $\omega t'-\varphi$. It means that change of $\omega t'$ is equivalent to change of $\varphi $. In other words, geometry of the electric and magnetic field looks like as the field rotates with angular velocity $\omega $ around $z$-axis. This conclusion relates also  to geometry of the field lines of  electric and magnetic fields. At first glance it would seem that we have a paradox: the linear velocity of motion is getting more then speed of light well away from $z$-axis.
 But motion of the field lines  does not relate to transfer of matter or field energy. The above equations state  only that the electromagnetic field at any  point ($r,\theta ,\varphi$) of space is equal to value of the field at the point ($r,\theta ,\varphi-\delta\varphi$) at the moment  $t-\delta\varphi/\omega$. Evidently,  in the far field  zone only  radiation field remains, which moves radially with the speed of light.


\section{Effective potential energy
}
\subsection{Dynamics of a charged particle}
Let us find the Lagrangian for a charged particle in the field of precessing magnetic dipole \citep{Landau}:
\begin{equation}
\label{Lag17}
L=-mc^{2}\sqrt{1-\displaystyle\frac{v^{2}}{c^{2}}}+\displaystyle\frac{e}{c}\left( \bm{A}\bm{v}\right),
\end{equation}
where  $\bm v$ is the particle velocity vector, $\bm A$ is the vector potential. It is easy to prove   that the vector potential
\begin{equation}
\label{10}
\bm{A}=-\frac{(\bm{n}\times\bm{\mu})}{r^{2}}-\frac{(\bm{n}\times\dot{\bm{\mu}})}{rc}
\end{equation}
gives the field (\ref{1a}) and (\ref{1b}).
Substituting spherical components of vectors $\bm{A}$ and $\bm{v}$
\begin{eqnarray}
\label{14}
A_{r}&=&0,\\
\label{15}
A_{\theta}&=&\frac{\mu\sin\alpha}{r^{2}}(\sin\tau+\rho\cos\tau),\\
\label{16}
A_{\varphi}&=&\displaystyle\frac{\mu}{r^{2}}\left[\cos\alpha\sin\theta\right.\nonumber \\
& &\left.-\sin\alpha\cos\theta(\cos\tau-\rho\sin\tau)\right],\\
\bm{v}&=&(\dot{r},\,\dot{\theta}r,\,\dot{\varphi}r\sin\theta)
\end{eqnarray}
into Lagrangian (\ref{Lag17}), we obtain
\begin{eqnarray}
\label{18}
L&=&-mc^{2}\sqrt{1-\displaystyle\frac{\dot{r}^{2}+\dot{\theta}r^{2}+\dot{\varphi}^{2}r^{2}\sin\theta^{2}}{c^{2}}}\nonumber \\
& &+\displaystyle\frac{e\mu}{cr}\left\{\dot{\theta}\sin\alpha(\sin\tau+\rho\cos\tau)+\dot{\varphi}\sin^2\theta\cos\alpha\right.\nonumber \\
& &\left.-\dot{\varphi}\sin\theta\cos\theta\sin\alpha(\cos\tau-\rho\sin\tau)\right\}.
\end{eqnarray}
This Lagrangian is a function of all spherical coordinates and  time. The time dependence appears in this formula as  combination $\omega t'-\varphi$.

Substituting this function into Lagrange's equation
\begin{equation}
\frac{d}{dt}\frac{\partial L}{\partial \dot q}-\frac{\partial L}{\partial q}=0,
\end{equation}
we obtain the relativistic equations of motion of a charged particle in the field of precessing magnetic dipole (see Appendix \ref{app1}).

Further we consider the charged particle to be  a nonrelativistic one.
Let us introduce a new set of generalized coordinates $\rho,\theta,\psi$ with
\begin{eqnarray}
\label{29}
\rho=\frac{r\omega}{c},\quad \psi=\varphi-\omega t.
\end{eqnarray}
Actually, this means that we use a co-rotating reference system.  It is known that the rotating reference system is applicable only inside the light cylinder, i.e. $r\omega<c$ or $\rho<1$. Then the nonrelativistic Lagrangian function takes the form
 \begin{eqnarray}
\label{Lagran2}
L&=&\frac{mc^2}{2\omega^2}\left[\dot\rho^2+\rho^2\dot{\theta}^2+\rho^2(\dot{\psi}+\omega)^{2}\sin^2\theta\right]\\
&&-\displaystyle\frac{e\mu\omega}{c^2\rho}\sin\alpha
\left[(\dot\psi+\omega)\sin\theta\cos\theta(\cos\xi+\rho\sin\xi) \right.\nn\\
&&+\left.\dot{\theta}(\sin\xi-\rho\cos\xi)\right]
+\frac{e\mu\omega}{c^2\rho}\cos\alpha
(\dot{\psi}+\omega)\sin^2\theta,\nonumber
\end{eqnarray}
where $\xi=\psi+\rho$. Now, the Lagrangian and Hamiltonian functions do not depend on time explicitly. In this case the value of the Hamiltonian is an integral of motion
\begin{eqnarray}
\label{Hamilton}
{\cal H}&=&q_i\frac{\partial L}{\partial \dot q_i}-L=\frac{mc^2}{2\omega^2}(\dot{\rho}^2+\rho^2\dot{\theta}^2+\rho^2\sin^2\theta\dot{\psi}^2)\nonumber \\
& &-\frac12 mc^2\rho^2\sin^2\theta-\frac{e\mu\omega^2}{c^2\rho}\cos\alpha\sin^2\theta\nonumber \\
& &+\frac{e\mu\omega^2\sin 2\theta}{2c^2\rho}\sin\alpha(\cos \xi+\rho\sin\xi).
\end{eqnarray}
The first term in this expression $K= \displaystyle\frac{mc^2}{2\omega^2}(\dot{\rho}^2+\rho^2\dot{\theta}^2+\rho^2\sin^2\theta\dot{\psi}^2)$  is always positive and it can be considered as the kinetic energy. The rest terms are usually  referred to as  effective potential energy. It can be written as:
\begin{eqnarray}
\label{32a}
V_{ef}&=&\displaystyle\frac{mc^2}{2}\left\{-\rho^2\sin^2\theta+\displaystyle\frac{N_{\perp}\sin2\theta}{\rho}(\cos\xi+\rho\sin\xi)\right.\nonumber \\
& &\left.-\displaystyle\frac{2N_{\parallel}\sin^2\theta}{\rho}\right\},
\end{eqnarray}
$$N_{\perp}=\displaystyle\frac{e\mu\omega^2\sin\alpha}{m c^4},\quad N_{\parallel}=\displaystyle\frac{e\mu\omega^2\cos\alpha}{m c^4}.$$

In this notation formula (\ref{Hamilton})  can be  represented as:
\begin{equation}\label{kinetik}
K={\cal H}- V_{ef}.
\end{equation}
The inequality ${\cal H}- V_{ef}\geq 0$ imposes restrictions on possible area of the particle motion.
\subsection{Stationary points of the effective potential energy}\label{points}
Let us investigate the effective potential energy. In the stationary points of the effective potential energy the particle can be in a stable, unstable or indifferent equilibrium.
The aim of this investigation is the stability of a charged particle at the stationary points.
As the Lagrangian is written for a non-relativistic particle, we consider that $\omega r\ll c$ or $\rho\ll 1$. This means that the particles moving around the axis of precession with an angular velocity of about $\omega$ are non-relativistic. In this approximation $\xi\approx\psi$ and the effective potential energy takes the form
\begin{eqnarray}
\label{32approx}
V_{ef}&=&\displaystyle\frac{mc^2}{2}\left\{-\rho^2\sin^2\theta+\displaystyle\frac{N_{\perp}\sin2\theta}{\rho}\cos\psi\right.\nonumber \\
& &\left.-\displaystyle\frac{2N_{\parallel}\sin^2\theta}{\rho}\right\}.
\end{eqnarray}
In order to obtain the stationary points of the function $V_{ef}$ we find  solutions of the set of equations:
\begin{eqnarray}
\label{32b}
\displaystyle\frac{\partial V_{ef}}{\partial q_{i}}=0,
\end{eqnarray}
where  $q_i=\rho,\theta,\psi$.
Hence, we have a system of three equations
\begin{eqnarray}
\label{33}
-2\rho^3\sin^2\theta+2 N_{\parallel}\sin^2\theta-N_{\perp}\sin2\theta\cos\psi=0,\\
\label{33a}
-\rho^3\sin2\theta-2 N_{\parallel}\sin2\theta+2N_{\perp}\cos2\theta\cos\psi=0,\\
\label{33b}
N_{\perp}\sin2\theta\sin\psi=0.
\end{eqnarray}
Equation (\ref{33b}) has next solutions:
 \begin{eqnarray}
\label{26aaa}
\psi=0,\,\pi,\\
\label{25b}
\theta=\frac{\pi n}{2},\quad (n\in Z).
\end{eqnarray}
\subsubsection{Solution for $\bm{\psi=0, \pi}$}
Using Eq. (\ref{26aaa}) we can eliminate the variable $\psi$ from the equations (\ref{33}) and  (\ref{33a}) by substitution $\cos\psi=\varepsilon$,
where $ \varepsilon=+1$ corresponds to $\psi=0$ and $\varepsilon=-1$ corresponds to $\psi=\pi$.
This results in a system of two equations with two unknowns:
\begin{eqnarray}
\label{256}
-\rho^3+N_{\parallel}-\varepsilon N_{\perp} \cot\theta=0\\
\label{256a}
-\rho^3\cot\theta-2N_{\parallel}\cot\theta+\varepsilon N_{\perp}(\cot^2\theta-1)=0
\end{eqnarray}
Expressing $ \cot\theta$ from equation (\ref{256}) and substituting it into equation (\ref{256a}), we obtain:
\begin{equation}
\label{eq-pho}
2\rho^6-N_{\parallel}\rho^3-N^2=0,
\end{equation}
where $N=\displaystyle\frac{e\mu\omega^2}{m c^4}$.
This equation has a solution:

\begin{eqnarray}\label{27}
\rho^3=\frac N4\left[ \cos\alpha\pm\sqrt{9-\sin^2\alpha}\right].
\end{eqnarray}

The sign before the square root is defined by the sign of the charge which is hidden in $N$, and by condition $\rho>0$.
Therefore, Eq. (\ref{27}) takes the form
\begin{equation}\label{27b}
\rho^3=\frac N4\left[ \cos\alpha+q\sqrt{9-\sin^2\alpha}\right],
\end{equation}
where $q=e/|e|=\pm 1$ is the sign of the charge.
Inserting $\rho$ into Eq. (\ref{256}), we find:
\begin{eqnarray}
\label{26a}
\tan\theta=- \displaystyle\frac{\varepsilon}{2\sin\alpha}\left[3\cos\alpha+q\sqrt{9-\sin^2\alpha}\right].
\end{eqnarray}

Thus, considering (\ref{26aaa}), (\ref{27}), (\ref{26a}),  for the positive particles in the case when $\sin2\theta\neq0$ we get the following series of coordinates of stationary points for $V_{ef}$:

\begin{eqnarray}
\label{27a}
& &\rho^3_{1}=\frac N4\left[ \cos\alpha+\sqrt{9-\sin^2\alpha}\right], \quad\psi_{1}=0, \nonumber \\
&  &\quad\tan\theta_{1}=-\displaystyle\frac{1}{2\sin\alpha}\left[3\cos\alpha+\sqrt{9-\sin^2\alpha}\right],
\end{eqnarray}
\begin{eqnarray}
\label{27s}
\rho^3_{2}=\displaystyle\frac{N}{4}\left[ \cos\alpha+\sqrt{9-\sin^2\alpha}\right],\quad \psi_{2}=\pi, \nonumber \\
\quad\tan\theta_{2}=\displaystyle\frac{1}{2\sin\alpha}\left[3\cos\alpha+\sqrt{9-\sin^2\alpha}\right].
\end{eqnarray}
For the negatively charged particles, we have next two stationary points:
\begin{eqnarray}
\label{27m}
\rho^3_{3}=\displaystyle\frac{N}{4}\left[ \cos\alpha-\sqrt{9-\sin^2\alpha}\right], \quad \psi_{3}=0,\nonumber \\
\quad\tan\theta_{3}=\displaystyle\frac{1}{2\sin\alpha}\left[-3\cos\alpha+\sqrt{9-\sin^2\alpha}\right],
\end{eqnarray}
\begin{eqnarray}
\label{27bb}
\rho^3_{4}=\displaystyle\frac{N}{4}\left[ \cos\alpha-\sqrt{9-\sin^2\alpha}\right], \quad\psi_{4}=\pi,\nonumber \\
\quad\tan\theta_{4}=\displaystyle\frac{1}{2\sin\alpha}\left[3\cos\alpha-\sqrt{9-\sin^2\alpha}\right].
\end{eqnarray}

Hence, the solution of equations (\ref{33}) -- (\ref{33b}) for the case $\sin2\theta\neq0$ gives two stationary points for a positive charge  and two points  for a negative charge.
\subsubsection{Solution for $\theta=\displaystyle\frac{\pi n}{2}$}
It follows from Eqs (\ref{33}) -- (\ref{33b}) that at the axis of rotation ($\theta=0,\,\pi$) all the first derivatives from effective potential energy are equal to zero only in the plane $\cos\psi=0$ and for any $\rho$. Which means that on the axis  $\theta=0,\,\pi$ there are not stationary points.

 As to the equatorial plane $\theta=\displaystyle\frac{\pi}{2}$,  Eqs (\ref{33}) -- (\ref{33b}) have the next solution:
\begin{eqnarray}
\label{N13}
\rho=N_{\parallel}^{\frac{1}{3}},
\quad
\cos\psi=0,\quad N_{\parallel}>0.
\end{eqnarray}
This gives two solutions for $\psi$, because  $\psi\in(0,2\pi)$.
Thus, in case $\theta=\displaystyle\frac{\pi}{2}$ there are two stationary points:
\begin{eqnarray}
\label{40a}
\rho_{5}=N_{\parallel}^{\frac{1}{3}}, \quad \theta_5=\displaystyle\frac{\pi}{2},\quad\psi_5=\displaystyle\frac{\pi}{2},\\
\label{40b}
\rho_{6}=N_{\parallel}^{\frac{1}{3}},\quad\theta_6=\displaystyle\frac{\pi}{2},\quad \psi_6=\displaystyle\frac{3 \pi}{2}.
\end{eqnarray}
It follows from (\ref{N13}) that $e\cos\alpha>0$, which means that the two above mentioned stationary points correspond to a positive charge if $\alpha <\pi/2$ and to a negative charge if
$\alpha >\pi/2$.

Particle located at a stationary point can be in a state of equilibrium.
Let us verify whether a particle with initial coordinates  $(\rho_i,\theta_i,\psi_i)$ defined by Eqs (\ref{27a}) -- (\ref{40b}) and initial zero velocity with respect to the rotating reference frame will be in equilibrium position. Substituting these coordinates  and $\dot\rho=\dot\theta=\dot\psi=0$  in equations of motion (\ref{50}) -- (\ref{52}) and taking into account that $\varphi=\omega t+\psi$ we  obtain identical equalities. Hence, a particle being at rest at one of the stationary points in the co-rotating coordinate system will keep this position.
 This means that in laboratory reference frame the particle is moving along a circle with constant velocity $v_i=c\rho_i\sin\theta_i$. Thus, there are at least six particular solutions  which describe circular motion of the particles in the field of inclined rotating dipole.
Positions of the orbits defined by angle $\theta$ and their radius  are different for the positive and negative particle.
Two of the trajectories are laying in the equatorial plane $z=0$.

\subsection{Labelling the stationary points}

In order to define whether the stationary points (\ref{27a}) -- (\ref{40b}) represent  maximum or minimum of potential energy we apply Sylvester's criterion: suppose that in some neighbourhood of a stationary point $M_{i}(\rho_{i},\theta_{i}, \psi_{i})$ the second derivatives  of the function $V_{ef}(\rho, \theta, \psi)$ are defined. The function has a local  maximum if the quadratic form  $d^{2}V_{ef}(M_{0})$ is negative definite and a local minimum if $d^{2}V_{ef}(M_{i})$ is positive definite. If the form $d^{2}f(M_{i})$ is alternating  then   there is no extremum at the point $M_{i}$ \citep{Merkin}.
Let us find all second partial derivatives and define their values at the stationary points (\ref{27a}) -- (\ref{27bb}).
\begin{eqnarray}
\label{35}
a_{11}&=&\frac{2}{mc^2}\frac{\partial^{2} V}{\partial\rho^2}=\cos^2\alpha-4\nn\\
&&-q\cos\alpha\sqrt{9-\sin^2\alpha},\\
\label{36}
a_{22}&=&\displaystyle\frac{2}{mc^2}\displaystyle\frac{\partial^{2} V}{\partial\theta^2}=6\rho^2,\\
\label{36-1}
a_{33}&=&\displaystyle\frac{2}{mc^2}\displaystyle\frac{\partial^{2} V}{\partial \psi^2}=-\displaystyle\frac{N\sin^2\alpha}{3\rho}(\cos\alpha\nn\\
&&-q\sqrt{9-\sin^2\alpha}),\\
\label{37}
a_{12}&=&\displaystyle\frac{2}{mc^2}\displaystyle\frac{\partial^{2} V}{\partial \rho\partial\theta}=\displaystyle\frac{2N\cos\psi\sin\alpha}{\rho^2},\\
\label{38}
a_{23}&=&\displaystyle\frac{2}{mc^2}\displaystyle\frac{\partial^{2} V}{\partial\theta\partial \psi}=0,\\
\label{39}
a_{13}&=&\displaystyle\frac{2}{mc^2}\displaystyle\frac{\partial^{2} V}{\partial\rho\partial \psi}=0.
\end{eqnarray}
Using Eq. (\ref{26a}) we transform  Eq. (\ref{35}) into
$a_{11}=-6\sin^2\theta$
and hence, it takes only negative values.

For the first fundamental form, we have:
\begin{eqnarray}
\label{36-2}
D_1&=&\left|\begin{array}{ccc} a_{11}& a_{12} \\  a_{21} &
a_{22} \\\end{array} \right|=
\frac{2N}{\rho}\left[-8\cos\alpha-\cos^3\alpha\right.\nonumber \\
& &+\left.(\cos^2\alpha-4)q\sqrt{9-\sin^2\alpha}\right].
\end{eqnarray}
The last equation can be transformed to
$$D_1=-\frac{2Nq}{\rho}\sin^2\theta\sqrt{9-\sin^2\alpha},$$
which makes obvious that $D_1<0$ both for negative and positive particles.
The third order determinant
\begin{eqnarray}
\label{37-1}
 \left|\begin{array}{ccc} a_{11}& a_{12}& a_{13} \\  a_{21} &
a_{22}&a_{23} \\ a_{31}&a_{32}&a_{33}\end{array} \right|=D_1a_{33}
\end{eqnarray}
is definitely negative because $a_{33}>0$.
Thus, the form $d^{2}f(M_{i})$ is alternating and there are no extremums of the function $V_{ef}$ at the  four points with coordinates (\ref{27a}) -- (\ref{27bb}) .

Calculation of the second derivatives in case of $\theta=\displaystyle\frac{\pi}{2}$ gives:
\bea
&& a_{11}=-6,\quad a_{22}=6N_{\parallel}^{\frac 23}, \quad a_{33}=0,\nn\\
&& a_{12}=0,  \quad a_{13}=0,  \quad a_{23}=2N_\perp N_\parallel^{-\frac 13}\sin\psi.\nn
 \eea
   The quadratic form combined of these derivatives is indefinite in its sign.

Summing up, we state that the six stationary points  (\ref{27a}) -- (\ref{27bb}), (\ref{40a}) and  (\ref{40b}) are points of equilibrium for a particle at rest, but the equilibrium is not stable. ''Particle at rest'' means that coordinates  $\rho,\theta,\psi$ are fixed. In the inertial reference frame the particle is moving in a circle with constant velocity.

Stationary points that we considered above  do not represent all solutions of the equations of motion for circular orbits. If we substitute into equations of motion (\ref{50}) -- (\ref{52}) $\rho=const$ we find one more solution for $e\cos\alpha>0$:
\begin{eqnarray}
\label{235a}
\theta=\frac{\pi}{2},\quad
\rho^3=2N,\quad\varphi=\Omega t+\varphi_0,
\end{eqnarray}
 where $\Omega=(1/2)\omega$, and  $\varphi_0$ is an arbitrary constant. This solution is valid for  a positive charge if $\alpha <\pi/2$ and for a negative one if $\alpha >\pi/2$.
\section{Equipotential surfaces of the effective potential energy}

Let us return to the expression (\ref{kinetik}): $K={\cal H}- V_{ef}$. The integral of motion $\cal H$ is defined by the  initial coordinates and velocity of the particle. If these are specified, the particle can move only in  space area where $V_{ef}<\cal H$. In this section we present plots of surfaces $V_{ef}=\rm const$. These surfaces restrict the permissible area for particle motion.

As we found in the previous section, the stationary points are at distance $\rho$ of order $|N|^{1/3}$ from the coordinate origin. Hence, we introduce a variable $\tilde{\rho}=\rho |N|^{-1/3}$ in order to make $\tilde{\rho} \sim 1$, and a reduced potential energy $\tilde V=2V_{ef}/(mc^2N^{2/3})$
\begin{equation}\label{surf_j}
\tilde V= -\tilde{\rho}^2\sin^2\theta+\frac{q\sin\alpha\sin2\theta}{\tilde{\rho}}\cos\psi
-\frac{2q\cos\alpha\sin^2\theta}{\tilde{\rho}},
\end{equation}
where $q=e/|e|$. The regions accessible to the particle motion are defined by inequality
\begin{equation}
\label{VlessC}
\tilde{V}\leq C.
\end{equation}
 The constant $C$ is  related to the integral of motion $\cal H$  as  $C=2{\cal H}/(mc^2N^{2/3})$.
\subsection {Aligned rotator, $\alpha=0$}
It is useful to start drawing plots from the simple case of a stationary dipole $\alpha=0$ in order to show how the axial symmetric equipotential surfaces transform into antisymmetric ones. If $\alpha=0$, the equipotential surface for a given constant $C$  is defined by equation ($\theta\neq 0,\,\pi$):
\begin{equation}
\label{236}
\tilde{\rho}^{3}+\eta\tilde{\rho}+2q=0,
\end{equation}
where $\eta=C/\sin^2\theta$.

Let us consider a case of positively charged particle.
Eq. (\ref{236}) has not positive roots for $\rho$ if  $C\geq 0$ and inequality (\ref{VlessC}) is satisfied in all space. There is no inaccessible area for particle motion.

\begin{figure}[htbp]
\begin{center}
\includegraphics [scale=0.22]{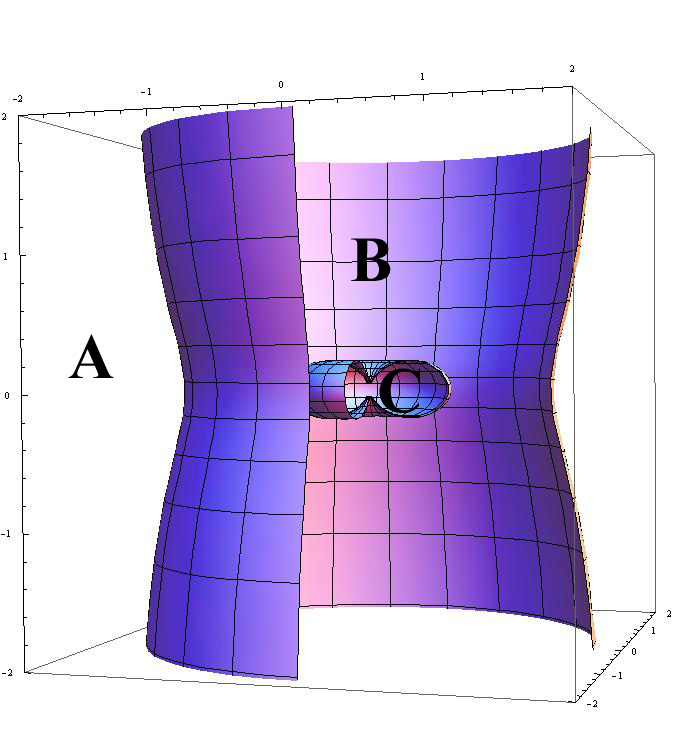}
\caption{Equipotential surfaces for $\alpha=0$, $C=-3.5$, $q=1$.
 C and A are accessible  regions, B --  inaccessible region}\label{fig3}
\end{center}
\end{figure}

\begin{figure}[htbp]
\begin{center}
\includegraphics [scale=0.22]{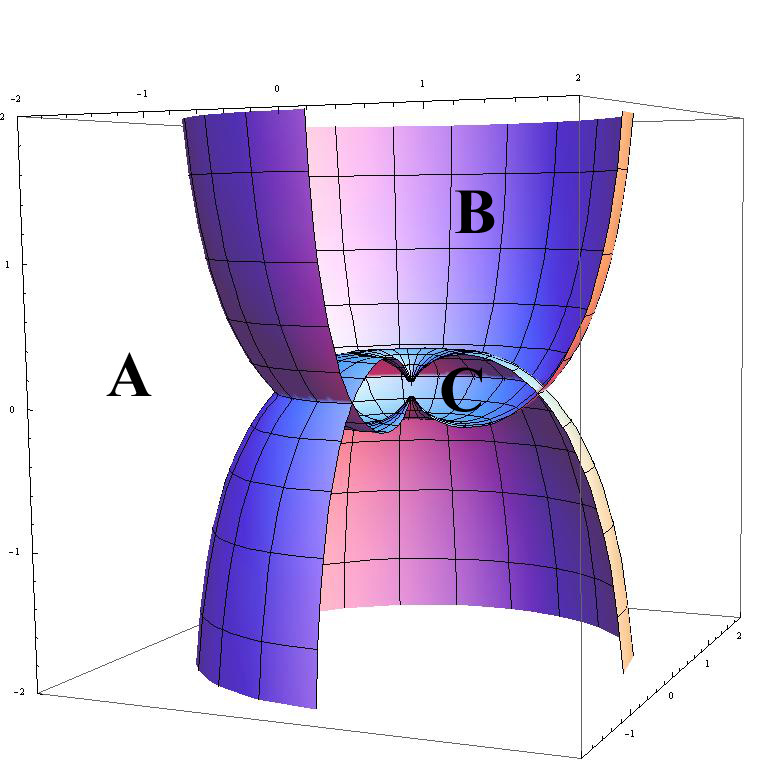}
\caption{Equipotential surface for $\alpha=0$, $C=-3$, $q=1$.
 A and C are accessible  regions, B --  inaccessible region.}\label{fig4}
\end{center}
\end{figure}

In case $C\in(-\infty;-3)$  Eq. (\ref{236}) has two roots at any $\theta$. The typical form of the equipotential surface in this case is given in Fig. \ref{fig3}. The inner surface has  form of a torus. Inequality  (\ref{VlessC}) is satisfied inside  the torus -- this is a region accessible for motion. The region between the torus and outer surface is inaccessible. Outside the external surface we have again accessible region.  As the constant $C$ increases from$-\infty$ to $-3$,   the inner surface of the torus is increasing, while the radius of outer surface is decreasing. At $C=-3$ the surfaces adjoin to each other along a circle of radius $\tilde\rho=1$. This  is shown in Fig.  \ref{fig4}.

\begin{figure}[htbp]
\center{
\includegraphics [scale=0.22]{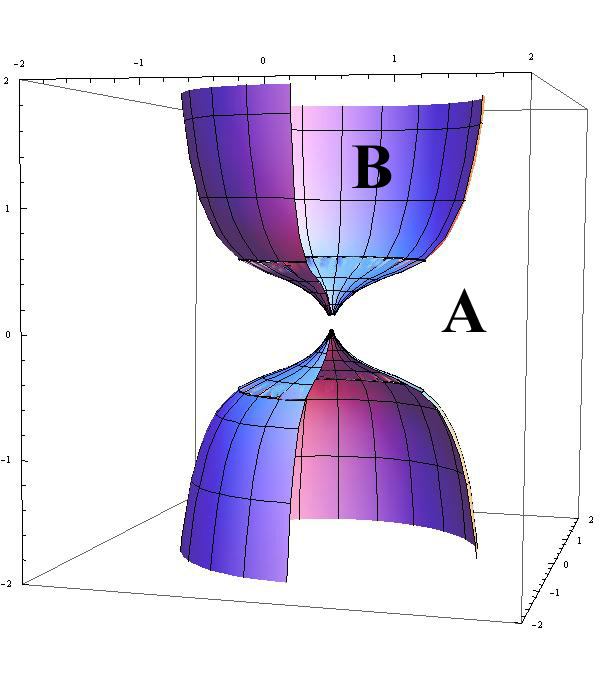}
}
\caption{Equipotential surface for $\alpha=0$, $C=-2$, $q=1$. B is inaccessible region, A -- accessible  region}
\label{fig2}\end{figure}

   In interval   $C\in(-3;0)$  Eq. (\ref{236}) has two positive roots  if $\sin^2\theta<-C/3$, one root if $\sin^2\theta=-C/3$ and no roots in case $\sin^2\theta>-C/3$ . The typical form of the equipotential surface in the case $C\in(-3;0)$ is given in Fig. \ref{fig2}. The space inside the surface is inaccessible for motion, while the outside space is accessible.
The upper and lower parts of the surface in Fig. \ref{fig2} have point-like connection at the coordinate origin $\tilde\rho=0$.  In the limit $C\to -0$ all the space is accessible for motion.

\begin{figure}[h!]
\begin{center}
\includegraphics [scale=0.22]{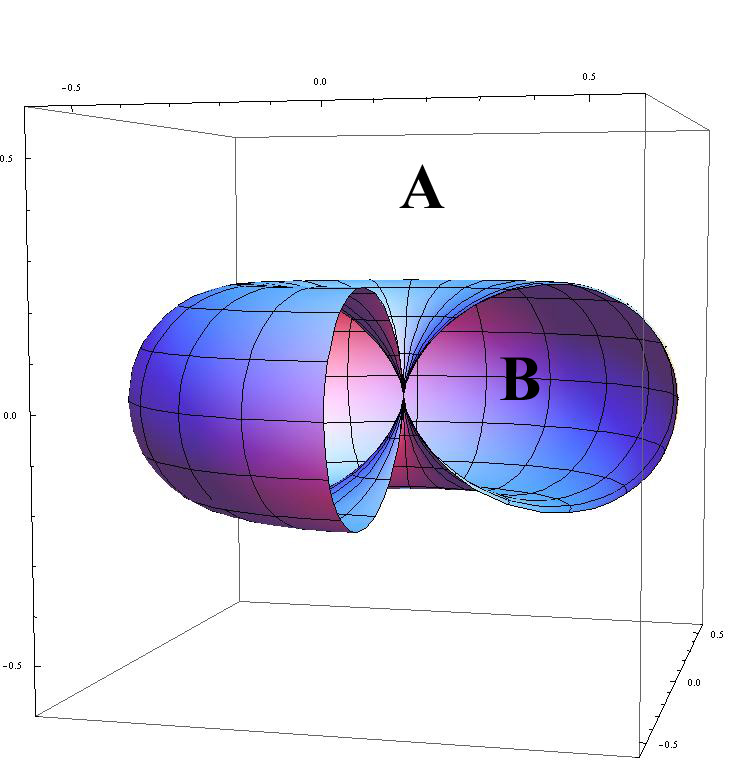}
\caption{Equipotential surfaces at $C=3, q=-1$, B - inaccessible region, A - accessible  region.}
\label{fig5}
\end{center}
\end{figure}

\begin{figure}[h!]
\begin{center}
\includegraphics [scale=0.22]{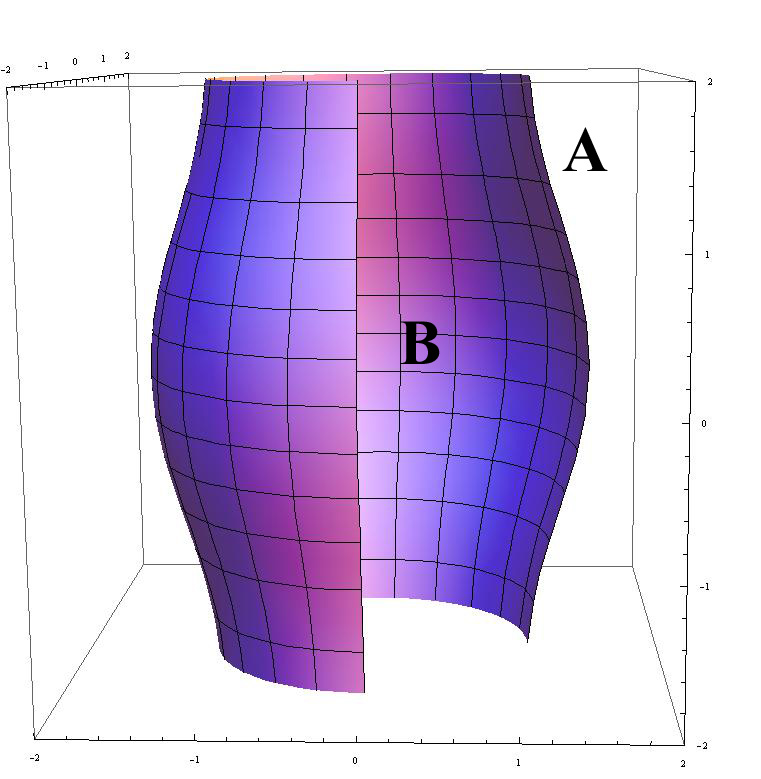}
\caption{Equipotential surface at $C=-2, q=-1$.
 A is accessible  region, B - inaccessible region.}
\label{fig6}
\end{center}
\end{figure}

For a negatively charged particle equation (\ref{236}) has one root for each $C$. In the case $C\in(0,+\infty)$ we have a surface in the form of a torus (Fig. \ref{fig5}). In other cases ($C\in(0,-\infty)$), the equipotential  surface has a form shown in Fig. \ref{fig6}. The radius of the surface increases with decreasing of $C$.

%
\subsection{General case,  $\alpha\neq 0$}
%

Let us consider the equipotential surfaces in the general case $\alpha\neq0$. One can see from equation (\ref{surf_j}) that the equipotential surfaces are symmetric with respect to substitution $\theta\to\pi-\theta$, $\psi\to\psi+\pi$, i.e. they are symmetric with respect to reflection in the plane $z=0$ and rotation around axis $OZ$ through $\pi$.

Instead of (\ref{236}) we have next equation
\begin{equation}
\label{nozero}
\tilde{\rho}^{3}+\eta\tilde{\rho}+2\chi=0,\quad \chi=q(\cos\alpha-\sin\alpha\cot\theta\cos\psi).
\end{equation}
In case $\chi<0$ this equation has one solution for $\tilde{\rho}$, hence, there is one equipotential surface which divides all space into accessible and inaccessible regions. If $\chi>0$ and $C>0$ the equation has  no solutions, all space is accessible for the particle motion.
 If $\chi>0$ and $C<0$, the number of solutions depends on the sign of expression $b=\chi- (-\eta/3)^{3/2}$. Namely, equation (\ref{nozero}) has no solutions for $\tilde\rho$ if $b>0$, one solution for  $b=0$ and two solutions if $b<0$.

The sign of $b$ is the same as the sign of expression
$$q\sin^2\theta(\cos\alpha\sin\theta-\sin\alpha\cos\theta\cos\psi)-\left(-\frac{C}{3}\right)^{-3/2}.$$
Evidently, the sum of first two terms lies in the range [-1, 1]. Hence, if $C<-3$, $b$ is definitely negative. In other cases sign of $b$ depends on $\theta$ and  $\alpha$.

\begin{figure}[h!]
\begin{center}
\includegraphics [scale=0.21]{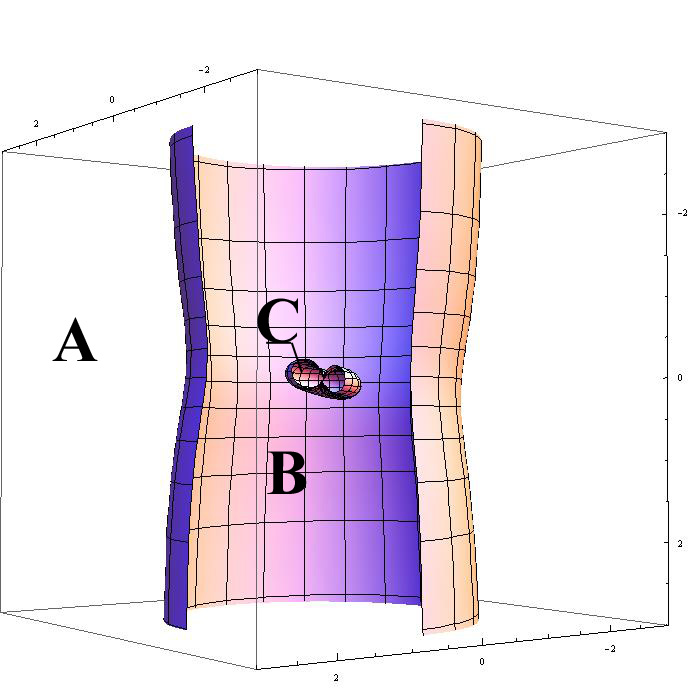}
\caption{Equipotential surfaces at $C=-4, q=1$.
 A and C  are accessible  regions, B - inaccessible region.}
\label{fig8}
\end{center}
\end{figure}

Thus,  if $C\in(-\infty;-3)$, equation (\ref{nozero}) has two roots at any $\theta$. The typical form of the equipotential surface in this case is given in Fig. \ref{fig8}. But unlike the case $\alpha=0$ (Fig. \ref{fig3}), there are three equipotential surfaces. 

\begin{figure}[h!]
\begin{center}
\includegraphics [scale=0.22]{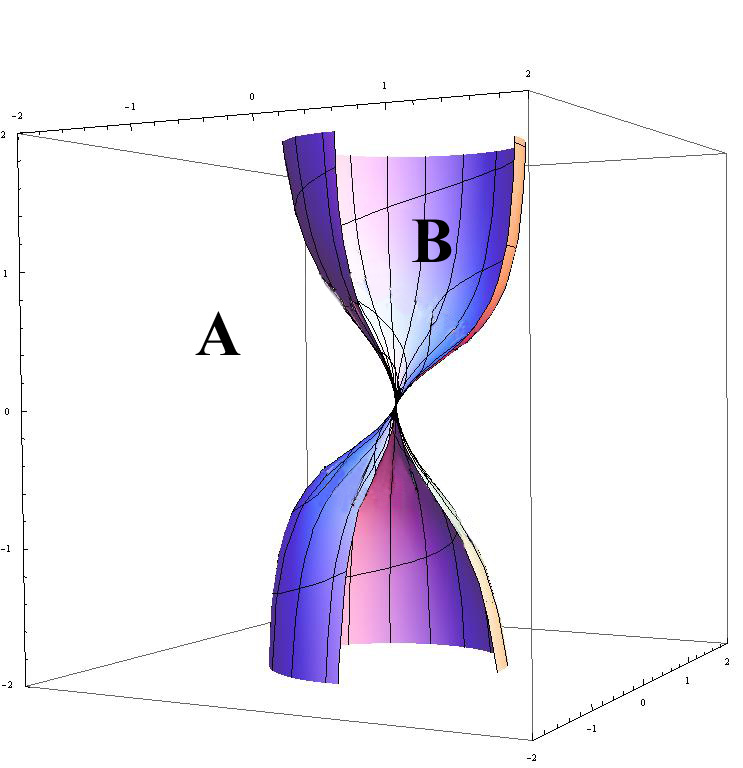}
\caption{Equipotential surface at $C=-1, q=1$.
A and C are accessible  regions, B - inaccessible region.}\label{fig7}
\end{center}
\end{figure}

\begin{figure}[h!]
\begin{center}
\includegraphics [scale=0.22]{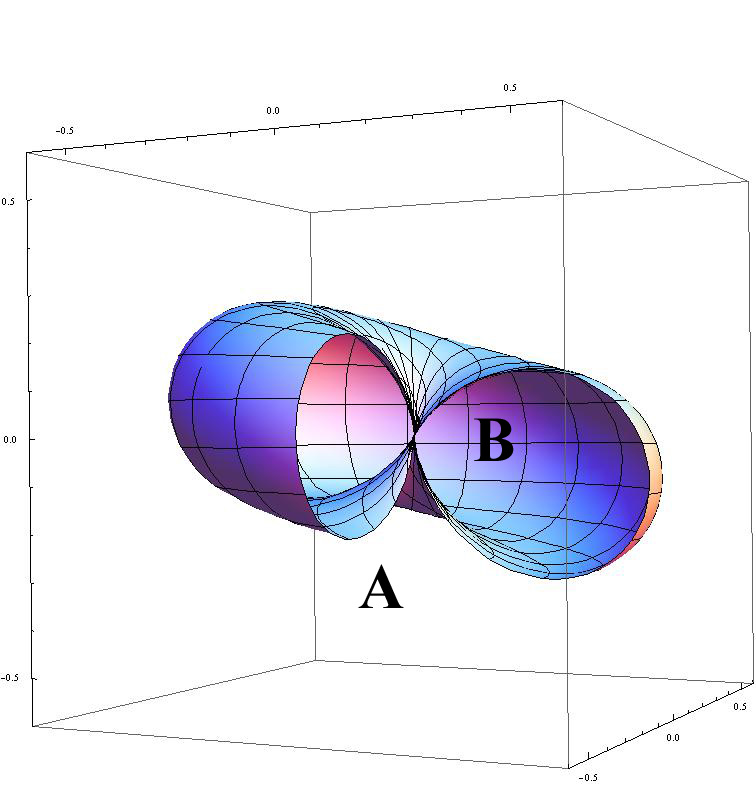}
\caption{Equipotential surface at $C=3, q=-1$.
  A is accessible  region,  B and C - inaccessible region.}
\label{fig9}
\end{center}
\end{figure}

\begin{figure}[h!]
\begin{center}
\includegraphics [scale=0.22]{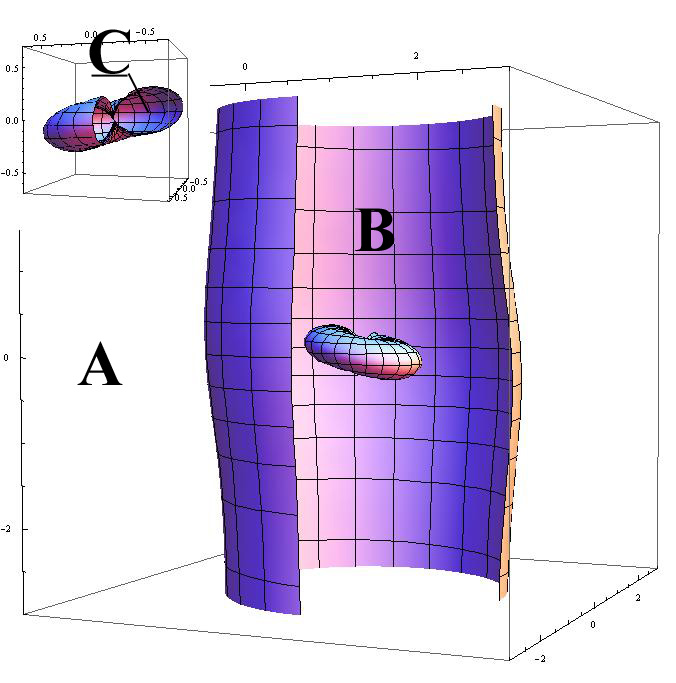}
\caption{Equipotential surfaces at $ C=-3, q=-1$. A and C are accessible  regions, B - inaccessible region. }
\label{fig10}
\end{center}
\end{figure}

 Equipotential surfaces for other values of $C$ and $q$ are shown in Figs \ref{fig7}-\ref{fig10}. All pictures \ref{fig8}-\ref{fig10} are plotted  for $\alpha=\pi/3$.
\subsection{Stationary points and equipotential surfaces}
Let us prove that the stationary points found in section \ref{points}
are located in accessible regions. One can see that in the limit $\alpha\to0$ all these points tend to points with coordinates  $\theta=\pi/2$,   $\rho=N^{1/3}, \psi=0, \pi, \pi/2, 3\pi/2$. A particle being at rest at these points in the rotating reference frame, is moving along a circle in the laboratory reference frame. Substituting the values of coordinates and velocity in Eq. (\ref{Hamilton}) we find that  $C=-3$. Hence, the particle moves along the border between two regions  A and C of  Fig.\ref{fig4}. This orbit  corresponds to unstable motion.

\begin{figure}[h!]
\begin{center}
\includegraphics [scale=0.22]{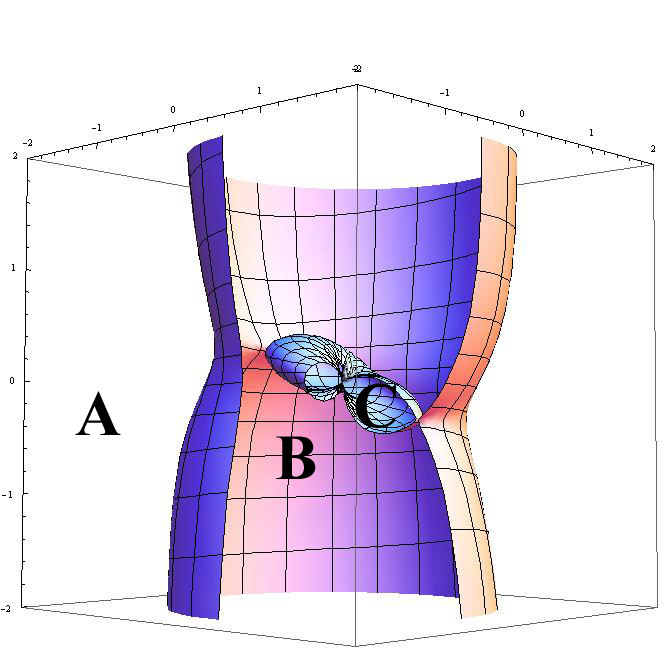}
\caption{Equipotential surfaces at $ C=-2.314, q=1.$  A and C are accessible  regions, B - inaccessible region.}
\label{fig10-1}
\end{center}
 \end{figure}

Let us find the constant $C$ for stationary points in case $\alpha\neq 0$. Substituting (\ref{27b}) and (\ref{26a}) in equation (\ref{surf_j})
we obtain:
\begin{eqnarray}
\label{237}
C=-\displaystyle\frac{\left(3\cos\alpha+q\sqrt{9-\sin^2\alpha}\right)}{2^{1/3}\left(\cos\alpha+q\sqrt{9-\sin^2\alpha}\right)^{1/3}}.
\end{eqnarray}
Then, for the stationary points (\ref{27a}), (\ref{27s}) we have $C\approx-2.314$. The equipotential surface for this value of $C$ is shown in Fig. \ref{fig10-1}. The stationary points are points of contiguity of regions $A$ and $C$. Again, a particle at these two positions is in unstable state of motion.

\begin{figure}[h!]
\begin{center}
\includegraphics [scale=0.22]{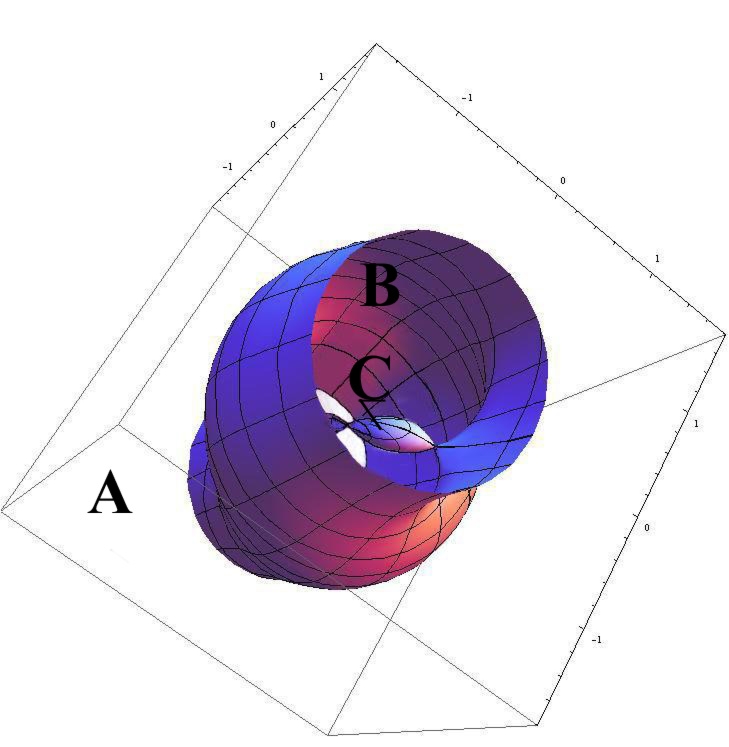}
\caption{Equipotential surfaces at $C=-0.816, q=-1.$ A and C are accessible  regions, B - inaccessible region.}
\label{fig11-1}
\end{center}
 \end{figure}

For the solutions (\ref{27m}), (\ref{27bb}) we have $C\approx-0.816$. This case is depicted in Fig. \ref{fig11-1}. We see that points  (\ref{27m}), (\ref{27bb}) are points of unstable motion.



\section{Conclusions}
We have calculated the components of the magnetic and electric fields of a rotating non-conducting body when the axis of the dipole magnetic field of the body is inclined to the rotation axis.

 In order to find whether the ''radiation belts'' are possible in such system, we have investigated the effective potential energy in the field of inclined rotating magnetic dipole. It is shown that there are closed equipotential surfaces which are co-rotating with the dipole field. The particles with definite initial energy are confined within such surfaces. The circular orbits of charged particles correspond to the points at which the two allowed for motion regions are contiguous to each other. It is shown that these orbits are unstable.

    Equipotential surfaces are plotted for different values of the constant of motion for positive and negative charge of a particle. The results can be used for description of the radiation belts around some specific celestial bodies possessing inclined dipolar magnetic field. 
\begin{acknowledgements}
This research has been supported by the grant for LRSS, project No 224.2012.2
\end{acknowledgements}

\appendix
\section{Relativistic equations of motion}\label{app1}
Here we record relativistic equations of motion of a charged particle  in the field of rotating magnetic dipole. Lagrangian (\ref{18}) leads to the following equations of motion in a spherical coordinate system.
\begin{eqnarray}
\label{50}
&&\frac{m}{(1-\beta^{2})^{3/2}}\left\{\frac{\ddot{r}}{c^2}(c^2-r^2\dot{\theta}^2-r^2\dot{\varphi}^2\sin^2\theta)+\frac{r\dot{r}}{c^2}(\dot{r}\dot{\theta}^2+r\dot{\theta}\ddot{\theta}+
\dot{\varphi}\sin^2\theta(\dot{r}\dot{\varphi}+r\ddot{\varphi})\right.\nonumber\\
&&\left.+\frac{1}{2}r\dot{\varphi}^2\dot{\theta}\sin2\theta)\right\}-
\frac{mr(\dot{\theta^{2}}+\dot{\varphi}^{2}\sin^2\theta)}{\sqrt{1-\beta^{2}}}+
\displaystyle\frac{e\mu\sin\alpha}{2cr^2}\left\{2\dot{\theta}(\sin\tau+\rho\cos\tau-\rho^{2}\sin\tau)\right.\nonumber \\
& &\left.-\dot{\varphi}\sin2\theta(\cos\tau-\rho\sin\tau-\rho^{2}\cos\tau)\right\}+\frac{e\mu\cos\alpha}{cr^2}\dot{\varphi}\sin^{2}\theta=0,
\end{eqnarray}
\begin{eqnarray}
\label{51}
&&\frac{m}{c^2(1-\beta^2)^{3/2}}\left\{r^2\ddot{\theta}(c^2-\dot{r}^2-r^2\dot{\varphi}^2\sin^2\theta)+r\dot{r}\dot{\theta}(2c^2-2\dot{r}^2-r^2\dot{\theta}^2-r^2\dot{\varphi}^2\sin^2\theta)\right.\nonumber \\
&&\left.+r^2\dot{\theta}(\dot{r}\ddot{r}+r^2\dot{\varphi}\ddot{\varphi}\sin^2\theta+\frac{1}{2}r^2\dot{\varphi}^2\dot{\theta}\sin2\theta)\right\}-\frac{m\dot{\varphi}^{2}r^{2}\cos\theta\sin\theta}{\sqrt{1-\beta^{2}}}\nn\\
&&-\displaystyle\frac{e\mu\sin\alpha}{r^2\omega}\left\{\rho(\omega-\dot{\rho} -2\dot{\varphi}\sin^2\theta)(\cos\tau-\rho\sin\tau)-\dot{\rho}\sin\tau\right\}
-\displaystyle\frac{e\mu\dot{\varphi}\cos\alpha\sin2\theta}{c r}=0,
\end{eqnarray}
\begin{eqnarray}
\label{52} &&\frac{m}{(1-\beta^2)^{3/2}}\left\{\sin^{2}\theta(2r\dot{r}\dot{\varphi}+r^2\ddot{\varphi})+r^2\dot{\varphi}\dot{\theta}\sin2\theta-\frac{r\sin^2\theta}{c^2}\left[\dot{r}\dot{\varphi}(2\dot{r}^2+r^2\dot{\theta}^2+
r^2\dot{\varphi}^2\sin^2\theta)\right.\right.\nonumber\\
& &\left.\left.+r\ddot{\varphi}(\dot{r}^2+r^2\dot{\theta}^2)-r\dot{\varphi}(\dot{r}\ddot{r}+r^2\dot{\theta}\ddot{\theta}+
\frac{1}{2}r^2\dot{\varphi}^2\dot{\theta}\sin2\theta)\right]-
\frac{r^2\dot{\varphi}\dot{\theta}}{c^2}\sin2\theta(\dot{r}^2+r^2\dot{\theta}^2+r^2\dot{\varphi}^2\sin^2\theta)\right\}\nonumber \\
& &+\frac{\mu e \sin\alpha}{r^2}\left\{\frac{2r\dot{\theta}\sin^2\theta}{c}(\cos\tau-\rho\sin\tau)+\frac{\dot{r}\sin2\theta}{2c}(\cos\tau-\rho\sin\tau-\rho^2\cos\tau)\right.\nonumber\\
& &\left.+\frac{\rho\sin 2\theta}{2}(\sin\tau+\rho\cos\tau)\right\}+\frac{\mu e\cos\alpha}{cr^2}(\dot{\theta}r\sin2\theta-\dot{r}\sin^2\theta) =0.
\end{eqnarray}

\section{Stability of the orbit in the equatorial plane}\label{appl2}
It is easy to check that equations of motion (\ref{50}) -- (\ref{52}) have partial solution
\begin{equation}
r=R,\quad \theta=\frac{\pi}{2},\quad
\label{40}
\varphi=\omega t+\varphi_0
\end{equation}
with restiction on the initial phase $\varphi_0$:
\[\cos(\rho_0+\varphi_0)+\rho_0\sin(\rho_0+\varphi_0)=0,\quad\rho_0=R\omega/c.\]
We assume now that the particle is a non relativistic one. Then $\rho_0\ll 1$ and we have for the initial phase $\cos\varphi_0=0.$
This solution  corresponds to motion of a particle being at rest at the stationary points (\ref{40a}) and (\ref{40b}).
Let us verify stability of this solution.
We develop equations of motion as series in increments of coordinates  in a neighbourhood of the solution. Let us introduce the  increments of coordinates as follows
\begin{equation}
\theta=\frac{\pi}{2}+\delta\theta,\quad
\label{40-1}
\varphi=\omega t+\varphi_0+\delta\varphi,\quad
r=R+\delta r.
\end{equation}
We substitute these coordinates in equations  (\ref{50}) -- (\ref{52}) and expand the equations up to first order of coordinates increments. Taking into account  that for a non reativistic particle $\rho\ll 1$ 
 we obtain a system of differential equations that describe the motion of a particle in a vicinity of a circle:
\begin{eqnarray}
\delta\ddot{r}-3\delta r\omega^2-\omega R\delta\dot{\varphi}+\varepsilon\omega R \tg\alpha\delta\dot{\theta}=0,\nonumber\\
\label{44}
\delta\ddot{\theta}-\varepsilon\displaystyle\frac{\delta\dot{r}\omega \tg\alpha}{R}+3\omega^2\delta\theta-\varepsilon\omega^2
\tg\alpha\delta\varphi=0,\\
\delta\ddot{\varphi}+\displaystyle\frac{\delta\dot{r}\omega}{R}-\varepsilon\omega^2\tg\alpha\delta\theta=0,\nn
\end{eqnarray}
where $\varepsilon=\sin\varphi_0=\pm1$.
We investigate the trivial solution of this system for stability.
By Lyapunov's theorem the trivial solution is stable when all roots of characteristic first-approximation equation of the system have negative real parts \citep{Merkin}.
We write Eq. (\ref{44}) in the form:
\begin{eqnarray}
x''_{1}-3x_{1}-x'_{3}+Tx'_{2}=0,\nonumber\\
\label{228}
x''_{2}-Tx'_{1}+3x_{2}-Tx_{3}=0,\\
x''_{3}+x'_{1}-Tx_{2}=0.\nonumber
\end{eqnarray}
where $x_{1}=\displaystyle\frac{\delta r}{R}, x_{2}=\delta\theta, x_{3}=\delta\varphi, T=\epsilon \tg\alpha$. Prime denotes $\xi$ derivative , where $\xi=\omega t$.
Reduce the system (\ref{228}) to a normal form:
\begin{eqnarray}
\label{229}
\begin{array}{ll}
&x'_{1}-d_{1}=0\\
&x'_{2}-d_{2}=0\\
&x'_{3}-d_{3}=0\\
\end{array}\quad
\begin{array}{rr}
d'_{1}-3x_{1}-d_{3}+Td_{2}=0&\\
d'_{2}-Td_1+3x_2-Tx_3=0&\\
d'_{3}+d_1-Tx_2=0&
\end{array}
\end{eqnarray}
A set of solutions of the system has the form
 $d_1=a_1e^{i\nu \xi}, d_2=a_2e^{i\nu\xi}....x_3=a_6e^{i\nu\xi}$.
This gives  the characteristic equation:
 \begin{eqnarray}
\label{230}
k^3-k^2(1+T^2)-3k(1+T^2)-3T^2=0,
\end{eqnarray}
where $k=\nu^2$.

Important criteria that give necessary and sufficient conditions for all
 the roots of the characteristic polynomial (with real coefficients) to
lie in the left half of the complex plane are known as Routh-Hurwitz
criteria \citep{Gantmacher}.

Given a polynomial
 \begin{eqnarray}
\label{231}
P(\lambda)=\lambda^n+a_1\lambda^{n-1}+...+a_{n-1}\lambda+a_n,
\end{eqnarray}
where the coefficients $a_i$ are real constants, $ i=1,2,3...,n$. Define the $n$ Hurwitz matrices using the coefficients $a_i$ of the characteristic polynomial:
 \begin{eqnarray}
\label{231a}
H_1=(a_1),\, H_2=\left(\begin{array}{ccc} a_{1}& 1 \\  a_{3} &
a_{2} \end{array} \right) ,\, H_3=\left(\begin{array}{ccc} a_{1}& 1& 0 \\  a_{3} &
a_{2}& a_{1} \\ a_{5}&a_{4}&a_{3}\end{array} \right) \dots
\end{eqnarray}

All of the roots of the polynomial $P(\lambda)$ are negative or have negative real part if the determinants of all Hurwitz matrices are positive (Det$ H_j\geq0$, $j=1,2,\dots,n$).

For the polynomials of degree n=3, the Routh-Hurwitz  criteria simplify to
 \begin{eqnarray}
\label{231-1}
a_1>0,\quad a_1a_2-a_3>0,\quad a_3(a_1a_2-a_3)>0.
\end{eqnarray}

From (\ref{230}) we get: $a_1=-(1+T^2),\, a_2=-3(1+T^2),\, a_3=-3T^2$. It is easy to see that  conditions (\ref{231-1}) of  Routh-Hurwitz
criteria  are not satisfied.
So, the trivial solution of system (\ref{44}) is not stable.

%


\begin{thebibliography}{}

\bibitem[Alfven(1950)]{Alfven} Alfven, H.  1950, ``Cosmical Electrodynamics'', International Series of Monographs on Physics,  Clarendon Press: Oxford
\bibitem[Babcock \& Cowling(1953)]{Babcock} Babcock, H. W., Cowling, T. G. 1953, \mnras, 113, 356
\bibitem[Cohen \& Kearney(1980)]{Cohen80}  Cohen, J. M. and Kearney, M. W. 1980
\apss 70, 295
\bibitem[Cohen \& Rosenblum(1972)]{Cohen} Cohen, J. M. and Rosenblum, A. 1972
\apss 16, 130
\bibitem[Deutsch(1955)]{Deutsch} Deutsch, A. J. 1955, Ann. d'Astrophys., 18, 1
\bibitem[DeVogelaere(1958)]{DeVogelaere} DeVogelaere, R., 1958,
In Contributions to the Theory of
Nonlinear Oscillations, ed. S. Lefschetz,  Princeton University Press: Princeton, p. 53-84
\bibitem[Dragt(1965)]{Dragt} Dragt, A. J. 1965,
Rev. Geophys., 3(2), 255
\bibitem[Epp \& Masterova(2012)]{EppMast} Epp, V., Masterova, M. A. 2012
Tomsk State Pedagogical University Bulletin 13, 51
\bibitem[Feynman(1964)]{Feynman} Feynman, R.P., Leighton,  R.B., Sands, M. 1964,  ``The Feynman Lectures on Physics''. Vol. 2,  Addison-Wesley: Reading
\bibitem[Gantmacher(1959)]{Gantmacher} Gantmacher, F.R. 1959, ``Applications of the Theory of Matrices'', Interscience Publishers Inc.: NY
\bibitem[Goldreich, P. \& Julian(1969) ]{Gold}Goldreich, P. and Julian, W. J. 1969 \apj  157, 869.
\bibitem[Holmes-Siedle \& Adams(2002)]{Holmes} Holmes-Siedle, A. G., Adams, L. 2002,``Handbook of Radiation Effects'', Oxford University Press: Oxford
\bibitem[Kaburaki(1981)]{Kaburaki} Kaburaki, O. 1981
\apss 74, 333
\bibitem[Katsiaris \& Psillakis(1986)]{Katsiaris} Katsiaris, G. A., Psillakis, Z. M. 1986
\apss 126, 69
\bibitem[Landau \&  Lifshitz(1975)]{Landau} Landau, L.D., Lifshitz, E. M. 1975, ``The Classical Theory of Fields'', 4th ed., Pergamon: NY
\bibitem[Lanzano(1968)]{Lanzano} Lanzano, P. 1968,
 \apss, 2, 319

\bibitem[Merkin(1996)]{Merkin} Merkin, R. M. 1996, ``Introduction to the Theory of stability'', Springer: NY 
\bibitem[Michel(1991)]{Michel} Michel, F.C. 1991, ``Theory of Neutron Star Magnetospheres'',    The University of Chicago Press: Chicago and London
\bibitem[Sarychev(2009)]{Sarych} Sarychev, V. T. 2009,
 Radiophys. Quant. Electr.,  52, 900
\bibitem[St\o rmer(1907)]{Stormer}St\o rmer, C. 1907,
 Arch. Sci. Phys. Nat., 24, 5–18;
ibid 113-158;  ibid 221-247.
\end{thebibliography}
\end{document}